\documentclass[reprint]{revtex4-1}
\usepackage{dcolumn}
\usepackage{graphicx}
\usepackage{float}
\usepackage{physics}
\usepackage{ulem}
\usepackage[colorlinks=true,allcolors=blue]{hyperref}

\begin{document}

\title{Machine learning holographic black hole from lattice QCD equation of state}

\author{Xun Chen}
\email{chenxun@usc.edu.cn}
\affiliation{School of Nuclear Science and Technology, University of South China, Hengyang 421001, China}

\author{Mei Huang}
\email{huangmei@ucas.ac.cn; corresponding author}
\affiliation{School of Nuclear Science and Technology, University of Chinese Academy of Sciences, Beijing 100049, China}

\date{\today}

\begin{abstract}
 Based on lattice QCD results of equation of state (EOS) and baryon number susceptibility at zero baryon chemical potential, and supplemented by machine learning techniques, we construct the analytic form of the holographic black hole metric in the Einstein-Maxwell-Dilaton (EMD) framework for pure gluon, 2-flavor, and 2+1-flavor systems, respectively. The dilaton potentials solved from Einstein equations are in good agreement with the extended non-conformal DeWolfe-Gubser-Rosen (DGR) type dilaton potentials fixed by lattice QCD EOS, which indicates the robustness of the EMD framework. The predicted critical endpoint (CEP) in the 2+1-flavor system is located at $(T^c$=0.094GeV, $\mu^c_B$=0.74GeV), which is close to the results from the realistic Polyakov-Nambu-Jona-Lasinio(PNJL) model, the functional renormalization group, and the holographic model with extended DeWolfe-Gubser-Rosen dilaton potential.

\end{abstract}

\maketitle

{\it Introduction:} Exploring phase transitions and phase structures of Quantum Chromodynamics (QCD) matter under extreme conditions is essential for understanding phenomena in heavy ion collisions, the early universe, and neutron stars. It has been predicted that a critical endpoint (CEP) exists at a finite baryon chemical potential $\mu_B$ \cite{Pisarski:1983ms}, and it has attracted extensive attention for several decades both in theory and experiment\cite{Stephanov:1998dy,Hatta:2002sj,Stephanov:1999zu,Hatta:2003wn,Schwarz:1999dj,Zhuang:2000ub}.  Searching for the CEP has become one of the most important goals at high baryon densities in heavy ion collisions at relativistic heavy ion collision (RHIC) \cite{STAR:2010mib,STAR:2010vob,STAR:2013gus,Luo:2017faz,STAR:2020tga,STAR:2022etb}, as well as in future facilities, e.g., FAIR at Darmstadt, NICA in Dubna and HIAF in Huizhou.

Due to the sign problem, lattice QCD is not well adapted to finite chemical potential regions. The CEP has been extensively investigated in 4-dimension effective QCD models, e.g., the Nambu-Jona-Lasinio (NJL), linear sigma  model\cite{Nambu:1961tp,Nambu:1961fr}, and their Polyakov-loop extended version  \cite{McLerran:2008ua,Sasaki:2010jz,Li:2018ygx,Sun:2023kuu}, the Dyson-Schwinger equations (DSE) \cite{Gao:2016qkh,Qin:2010nq,Shi:2014zpa,Fischer:2014ata}, and
the functional renormalization group (FRG)\cite{Fu:2019hdw,Zhang:2017icm}. In recent decades, the holographic gauge-gravity duality \cite{Maldacena:1997re}  has been widely applied as an important nonperturbative method in describing hadron physics \cite{Erdmenger:2007cm,Brodsky:2014yha,Casalderrey-Solana:2011dxg,Adams:2012th} and QCD matter under extreme conditions  \cite{Gubser:2008ny,DeWolfe:2010he,Jarvinen:2022doa,He:2013qq,Yang:2014bqa,Yang:2015aia,Dudal:2017max,Dudal:2018ztm,Fang:2015ytf,Liu:2023pbt,Li:2022erd,Critelli:2017oub,Grefa:2021qvt,Arefeva:2020vae,Chen:2018vty,Chen:2020ath,Zhou:2020ssi,Chen:2019rez}. The 5-dimensional Einstein-Maxwell-Dilaton (EMD) framework \cite{He:2013qq,Yang:2014bqa,Yang:2015aia,Fang:2015ytf,Dudal:2017max,Dudal:2018ztm,Chen:2018vty,Chen:2020ath,Zhou:2020ssi,Chen:2019rez} has been adapted as the working framework for describing QCD matter at finite temperature and density.

A family of five-dimensional black holes dual to QCD equation of state has been constructed in Refs. \cite{Gubser:2008yx,Gubser:2008ny} with a non-conformal dilaton potential, and a CEP was firstly obtained from holographic dual black hole by DeWolfe-Gubser-Rosen (DGR) in \cite{DeWolfe:2010he}. Further careful studies have been conducted with extended DGR non-conformal dilaton potential with more parameters
\cite{Critelli:2017oub,Grefa:2021qvt,Hippert:2023ndj,Hippert:2023bel,Cai:2022omk,Zhao:2022uxc}, see review in \cite{Rougemont:2023gfz}.
Another equivalent method is the potential reconstruction method, where one can input the dilaton or a metric to determine the dilaton potential. Although this approach results in a temperature-dependent dilaton potential, the model can still capture many QCD properties through analytical solutions.

Machine learning has become a useful tool in high-energy physics; for a recent review, see Ref. \cite{Zhou:2023pti}. Furthermore, the integration of deep learning with holographic QCD has been explored in recent studies\cite{Hashimoto:2018ftp,Akutagawa:2020yeo,Hashimoto:2018bnb,Yan:2020wcd,Hashimoto:2021ihd,Song:2020agw,Ahn:2024gjf,Gu:2024lrz}. Unlike conventional holographic models, this approach first employs specific QCD data to determine the bulk metric (as well as other model parameters) through machine learning. Subsequently, the model utilizes the determined metric to calculate other physical QCD observables, serving as predictions of the model. This letter will offer an approach to construct a holographic model with the help of machine learning.

In this study, we will employ the potential reconstruction method, supplemented by machine learning, to extract the black hole metric of the EMD model from the lattice results of the EOS at zero chemical potential for pure gluon, 2-flavor, and 2+1-flavor systems, respectively. This model will then be used to predict the location of CEP.

{\it The general EMD framework:}
Firstly, we review the 5-dimensional Einstein-Maxwell-Dilaton systems \cite{He:2013qq,Yang:2014bqa,Yang:2015aia,Dudal:2017max,Dudal:2018ztm,Chen:2018vty,Chen:2020ath,Zhou:2020ssi,Chen:2019rez}. The action includes a gravity field $g_{\mu \nu}$, a Maxwell field $A_\mu$ and a dilaton field $\phi$. In the Einstein frame, it is expressed by the following equation:
\begin{equation}
\begin{aligned}
S_{E} & =\frac{1}{16 \pi G_5} \int d^5 x \sqrt{-g}\left[R-\frac{f(\phi)}{4}F^2-\frac{1}{2} \partial_\mu \phi \partial^\mu \phi-V(\phi)\right]. \\
\end{aligned}
\end{equation}
where $f(\phi)$ is the gauge kinetic function coupled with the gauge field $A_\mu$, $F$ is the tensor of Maxwell field, $V\left(\phi\right)$ is the dilaton potential and $G_5$ is the Newton constant in five dimensions. The explicit forms of the gauge kinetic function $f\left(\phi\right)$ and the dilaton potential $V\left(\phi\right)$ can be solved consistently from the equations of motion(EOMs).

We give the following ansatz of metric
\begin{equation}
d s^2=\frac{L^2 e^{2 A(z)}}{z^2}\left[-g(z) d t^2+\frac{d z^2}{g(z)}+d \vec{x}^2\right],
\end{equation}
where $z$ is the 5th-dimensional holographic coordinate and the radial $L$ of $\rm AdS_5$ space is set to be one, i.e., $L=1$.
The boundary conditions are
\begin{equation}
A(0)=-\sqrt{\frac{1}{6}} \phi(0), \quad g(0)=1, \quad A_t(0)=\mu+\rho^{\prime} z^2+\cdots.
\end{equation}
$\mu$ can be regarded as baryon chemical potential and $\rho^{\prime}$ is proportional to the baryon number density. $\mu$ is related to the quark-number chemical potential $\mu = 3\mu_q$. The baryon number density can be calculated as\cite{Critelli:2017oub,Zhang:2022uin}
\begin{equation}
\begin{aligned}
\rho & =\left|\lim _{z \rightarrow 0} \frac{\partial \mathcal{L}}{\partial\left(\partial_z A_t\right)}\right| \\
& =-\frac{1}{16\pi G_5} \lim _{z \rightarrow 0}\left[\frac{\mathrm{e}^{A(z)}}{z} f(\phi) \frac{\mathrm{d}}{\mathrm{d} z} A_t(z)\right].
\end{aligned}
\end{equation}
$\mathcal{L}$ is the Lagrangian density in the Einstein frame.

To obtain the analytical solution, we assume the form of $f(\phi)$ and $A(z)$ with several parameters. From experience in \cite{Chen:2020ath}, we take the ansatz of the metric
\begin{equation}
A(z)= d \ln(a z^2 + 1) + d \ln(b z^4 + 1),
\label{eq:az}
\end{equation}
and the gauge kinetic function $f(z)$ is taken as
\begin{equation}
f(z)=e^{c z^2-A(z)+k}.
\label{eq:fz}
\end{equation}

Then, we can get
\begin{equation}
\begin{aligned}
g(z) & =1-\frac{1}{\int_0^{z h} d x x^3 e^{-3 A(x)}}\Big[\int_0^z d x x^3 e^{-3 A(x)}. \\
& +\frac{2 c \mu^2 e^k}{\left(1-e^{-c z_h^2}\right)^2} \operatorname{det} \mathcal{G}\Big],\\
\phi^{\prime}(z) & =\sqrt{6\left(A^{\prime 2}-A^{\prime \prime}-2 A^{\prime} / z\right)}, \\
A_t(z) & =\mu \frac{e^{-c z^2}-e^{-c z_h^2}}{1-e^{-c z_h^2}}, \\
V(z) & =-3 z^2 g e^{-2 A}\Big[A^{\prime \prime}+A^{\prime}\left(3 A^{\prime}-\frac{6}{z}+\frac{3 g^{\prime}}{2 g}\right). \\
& -\frac{1}{z}\left(-\frac{4}{z}+\frac{3 g^{\prime}}{2 g}\right)+\frac{g^{\prime \prime}}{6 g}\Big],
\end{aligned}
\end{equation}
where
\begin{equation}
\operatorname{det} \mathcal{G}=\left|\begin{array}{ll}
\int_0^{z_h} d y y^3 e^{-3 A(y)} & \int_0^{z_h} d y y^3 e^{-3 A(y)-c y^2} \\
\int_{z_h}^z d y y^3 e^{-3 A(y)} & \int_{z_h}^z d y y^3 e^{-3 A(y)-c y^2}
\end{array}\right|.
\end{equation}
The Hawking temperature and entropy of this black hole solution are given by,
\begin{equation}
\begin{aligned}
T & =\frac{z_h^3 e^{-3 A\left(z_h\right)}}{4 \pi \int_0^{z_h} d y y^3 e^{-3 A(y)}}\Big[1+ \\
&\frac{2 c \mu^2 e^k\left(e^{-c z_h^2} \int_0^{z_h} d y y^3 e^{-3 A(y)}-\int_0^{z_h} d y y^3 e^{-3 A(y)} e^{-c y^2}\right)}{(1-e^{-c z_h^2})^2} \Big],
\end{aligned}
\end{equation}
\begin{equation}
s=\frac{e^{3 A\left(z_h\right)}}{4 G_5 z_h^3}.
\end{equation}
After knowing the entropy, the free energy can be calculated as
\begin{equation}
\begin{aligned}
F&=-\int s d T-\rho d \mu
\end{aligned}
\end{equation}
The pressure is defined as $p=-F$. The energy density of the system can be derived as
\begin{equation}
\epsilon=-p+s T+\mu \rho.
\end{equation}
The second-order baryon number susceptibility is defined as
\begin{equation}
\chi_2^B=\frac{1}{T^2} \frac{\partial \rho}{\partial \mu}.
\end{equation}

{\it Machine learning the holographic metric:} There are three undetermined parameters $a,b,d$ in $A(z)$, see Eq.(\ref{eq:az}) and two parameters $c,k$ in $f(z)$, see Eq.(\ref{eq:fz}), together with the Newton constant $G_5$, the parameter space is six-dimensional. All of these parameters can be simultaneously constrained by machine learning the lattice QCD results on the EOS and the baryon number susceptibility at zero chemical potential.
The lattice QCD results on EOS at zero chemical potential are taken from Refs. \cite{Borsanyi:2012ve,Burger:2014xga,HotQCD:2014kol} for pure gluon, 2-flavor and 2+1-flavor systems, respectively.  The lattice results of baryon number susceptibility $\chi_2^{B}$ are taken from Refs. \cite{Datta:2016ukp,Bazavov:2017dus}.

\begin{figure}
    \centering
    \includegraphics[width=7cm]{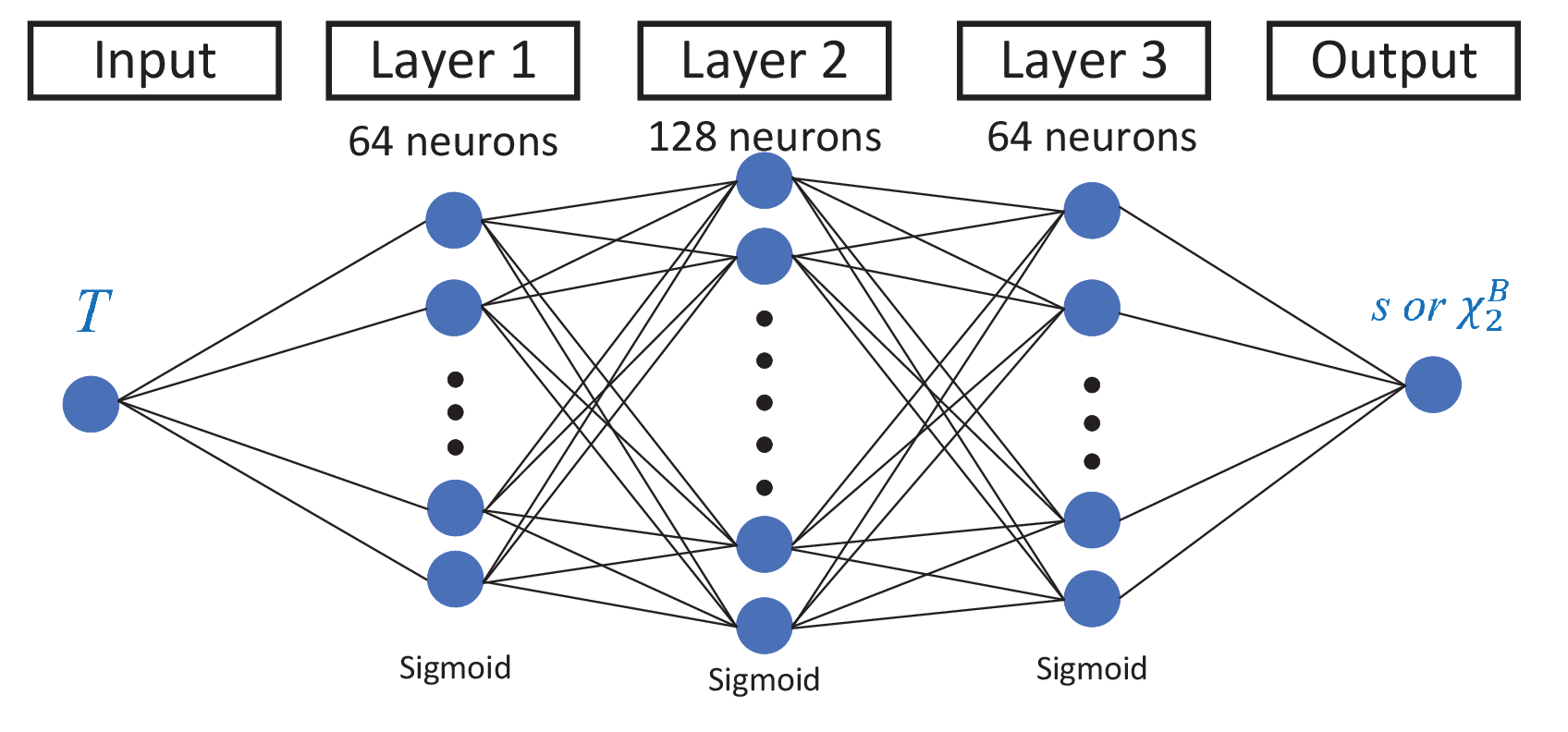}
    \caption{\label{sketch2} A sketch of the neural network used in our model. The input is the temperature $T$ and the output is the entropy $s$ or baryon number susceptibility $\chi_2^B$.}
\end{figure}

We implement a deep neural network for regression analysis using the TensorFlow framework as shown in Fig.\ref{sketch2}. The structure of the neural network consists of an input layer, three hidden layers with sigmoid activation functions, and a single-node output layer. The model employs mean squared error as the loss function and uses the Adam optimizer for parameter optimization. In our approach, we extract 35, 12, 55 points from the lattice's entropy measurements for pure gluon, 2 flavor, and 2+1 flavor systems. Additionally, we use 8 and 15 points from the lattice's baryon number susceptibility for 2 and 2+1 flavor systems respectively. The training process for our model encompasses 10,000 training epochs. After the training of the neural network model, we can obtain the relationship between the input variable, i.e., the temperature $T$, and the target output variable,i.e., the entropy $s$ or baryon number susceptibility $\chi_2^B$, so that the model can predict the target output as accurately as possible on the test set.

Now we turn to solving an optimization problem to find the optimal parameter values through a gradient descent algorithm. The program defines a loss function that employs the least squares method to measure the difference between the predictions of the model and the output from a pre-trained neural network. This comparison is utilized to evaluate the performance of model parameters $a$, $b$, $c$, $d$, $k$, and $G_5$. Initial values are assigned to these parameters, with constraints applied to $a \geq 0$ and $b \geq 0$ to maintain $A(z)$ within real number ranges. The optimization of parameters is conducted using the Adam optimizer through 5000 iterations of gradient descent, during which the loss is calculated and parameters are updated iteratively. Upon completion, it reports the final optimized values for all the parameters involved.

The machine learning process gives 6 optimized parameters as well as the predicted critical temperature $T_c$ at $\mu=0$ for pure gluon, 2-flavor and 2+1-flavor systems, respectively. The minimum of the speed of sound  determines $T_c = 0.265~\text{GeV}$ for pure gluon, $T_c = 0.189 \rm GeV$ for 2-flavor, and $T_c = 0.128 \rm GeV$ for 2+1-flavor system at vanishing chemical potential. The results are listed in Table \ref{table:parameter}.

\begin{table}[htbp]
	\centering
	\begin{tabular}{|c|c|c|c|c|c|c|c|}
		\hline  %
		\phantom&$a$&$b$&$c$&$d$&$k$&$G_5$&$T_c$ \\
		\hline
		$N_f = 0$&0$$&0.072$$&0$$&-0.584$$&0$$&1.326$$&0.265  \\
        \hline
		$N_f = 2$&0.067$$&0.023$$&-0.377$$&-0.382$$&0$$&0.885$$&0.189  \\
        \hline
        $N_f = 2+1$&0.204$$&0.013$$&-0.264$$&-0.173$$&-0.824$$&0.400$$&0.128   \\
        \hline
	\end{tabular}
\caption{Parameters given by the machine learning of pure gluon system, 2-flavor, and 2+1-flavor system, respectively. $T_c$ is the predicted critical temperature at vanishing chemical potential. The unit of $T$ is GeV.}  %
\label{table:parameter}
\end{table}
\begin{figure}
    \centering
    \includegraphics[width=7cm]{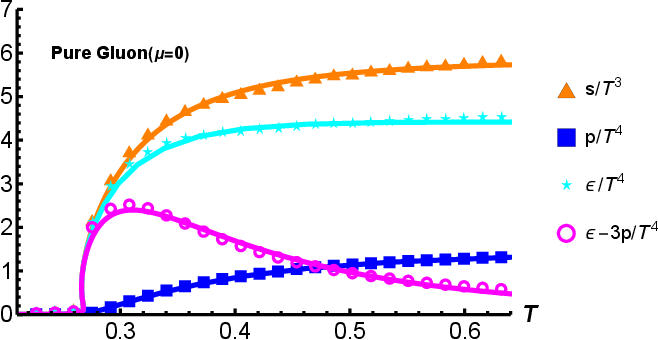}
    \caption{\label{pureeos}  The machine learning results (lines) of the entropy density, pressure, energy density and trace anomaly as a function of the temperature for pure gluon system at $\mu=0$, comparing with lattice results in different symbols with error bar\cite{Borsanyi:2012ve}. The unit of $T$ is GeV.}
\end{figure}
\begin{figure}
    \centering
    \includegraphics[width=7cm]{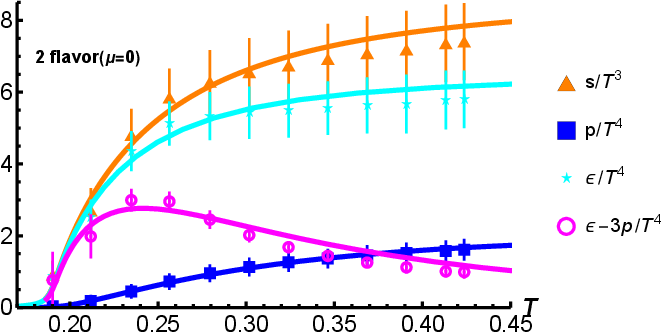}
    \caption{\label{2lattice} The machine learning results (lines) of
    the entropy density, pressure,  energy density and trace anomaly as a function of the temperature for the 2-flavor gluon system at $\mu=0$, comparing with lattice results in different symbols with error bar\cite{Burger:2014xga}. The unit of $T$ is GeV.}
\end{figure}
\begin{figure}
    \centering
    \includegraphics[width=7cm]{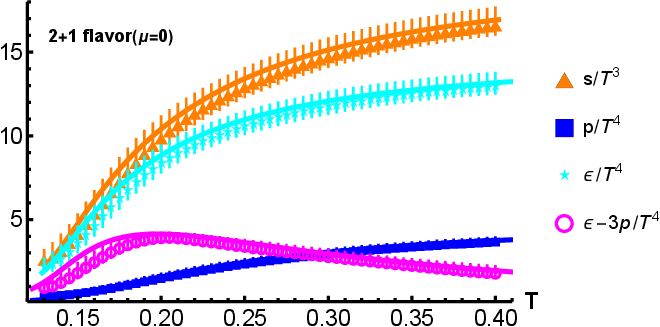}
    \caption{\label{2+1lattice} The machine learning results (lines) of
 the entropy density, pressure,  energy density and trace anomaly as a function of temperature for the 2+1-flavor system at $\mu=0$, comparing with lattice results in different symbols with error bar\cite{HotQCD:2014kol}. The unit of $T$ is GeV.}
\end{figure}

\begin{figure}
    \centering
    \includegraphics[width=9cm]{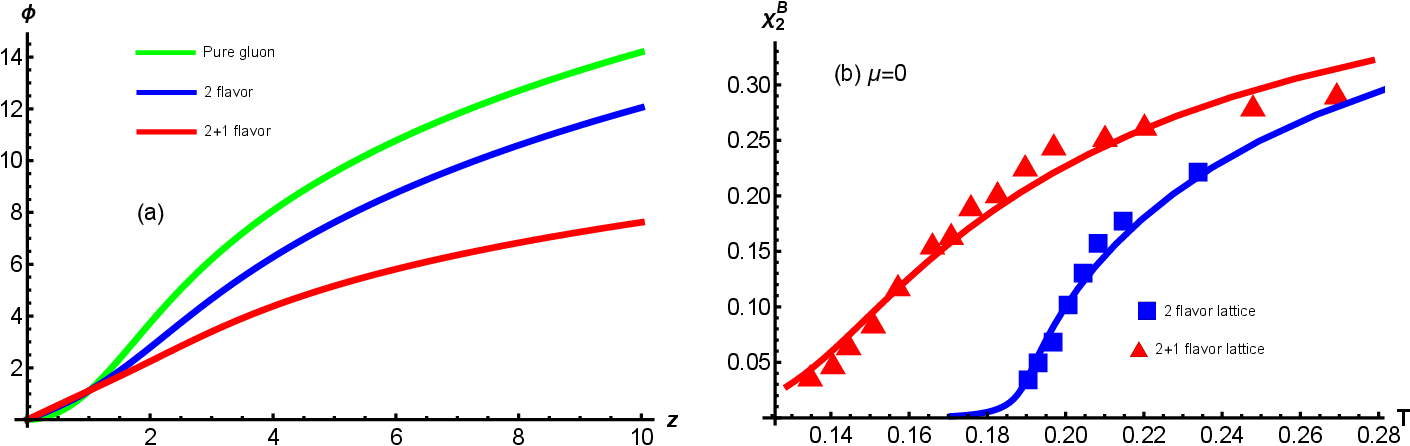}
    \caption{\label{comphi}  $\phi$(z) (a) and $\chi_2^B$ (b) for pure gluon(green), 2-flavor(blue) and 2+1-flavor (red) systems, respectively.  Lattice results come from Refs. \cite{Datta:2016ukp,Bazavov:2017dus}. The unit of $z$ is $\rm GeV^{-1}$. The unit of $T$ is GeV.}
\end{figure}

The machine learning results, in comparison with lattice results for the entropy density, pressure, energy density, and trace anomaly as functions of temperature, are shown for pure gluon, 2-flavor, and 2+1-flavor systems in Figs.\ref{pureeos}, \ref{2lattice}, and \ref{2+1lattice}. The results of $\phi$(z) and the baryon number susceptibility $\chi_2^B$ calculated from machine leaning $A(z)$and $f(z)$ are shown in Fig.\ref{comphi}. It shows that the results of $\chi_2^B$ are in good agreement with lattice results for 2-flavor and 2+1-flavor systems at $\mu=0$ around the phase transition temperature.

\begin{figure}
    \centering
    \includegraphics[width=9cm]{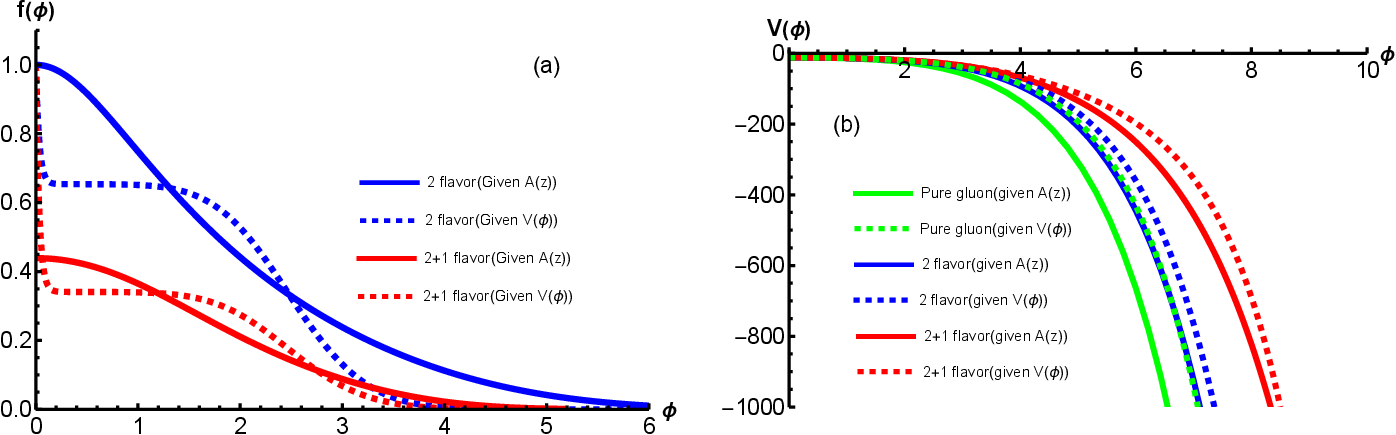}
    \caption{\label{comhphi}  $f(\phi)$ (a) and $V(\phi)$ (b) for pure gluon (green), 2-flavor(blue) and 2+1-flavor (red) systems, respectively. Solid lines represent the results obtained through machine learning, while dashed lines depict the outcomes from the extended DGR model in Ref. \cite{Zhao:2023gur}. }
\end{figure}
\begin{figure}
    \centering
    \includegraphics[width=8cm]{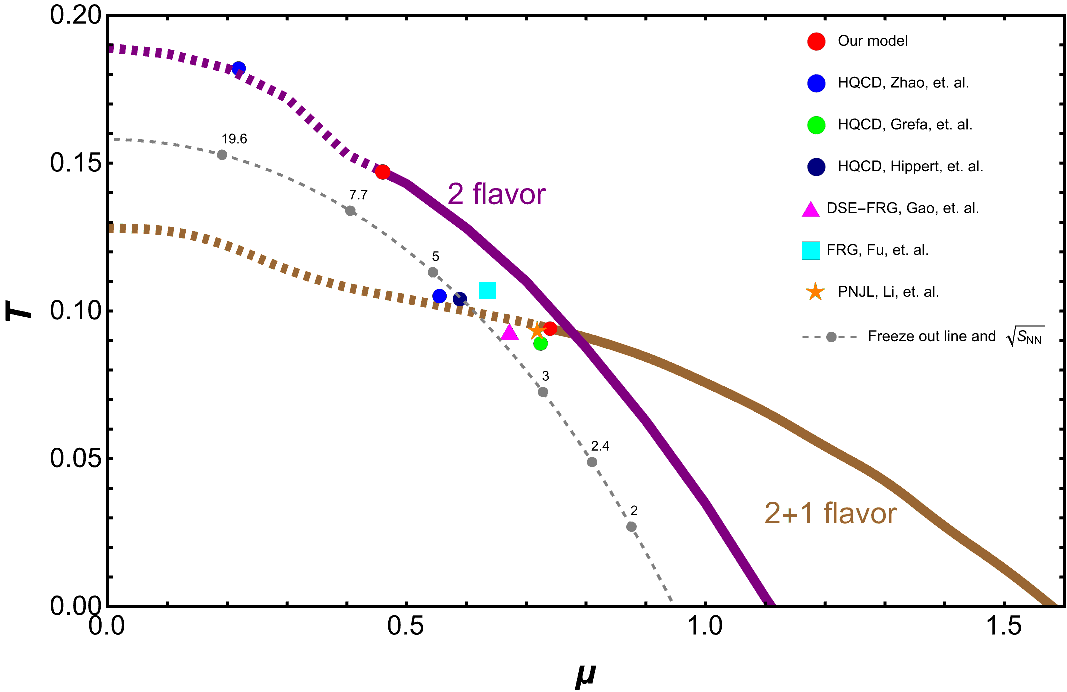}
    \caption{\label{cep} The location of CEP for 2 flavor and 2+1-flavor from theory predictions.  Red dots are our results,  blue dots \cite{Zhao:2022uxc}, dark-blue dots\cite{Hippert:2023bel}, and green dots \cite{Grefa:2021qvt} are from DGR-type holographic QCD models. The pink triangles, the blue squares, and the orange stars are from DSE-FRG \cite{Gao:2020qsj},  FRG \cite{Fu:2019hdw} and realistic PNJL model\cite{Li:2018ygx}, respectively. The grey dashed line is the freeze-out line $T(\mu)=0.158-0.14\mu^2-0.04\mu^4$ and the grey-dots are corresponding $\sqrt{s}$ with $\mu_{B}(\sqrt{s})=1.477/(1+0.343\sqrt{s})$ alone the freeze-out line\cite{Luo:2017faz}. }
\end{figure}

{\it Comparing with extended DGR models:}
In our framework, with given $A(z)$ and $f(z)$ from machine learning, we can easily solve the dilaton field $\phi$ and dilaton potential $V(\phi)$ as well as $f(\phi)$. As introduced in Introduction, by incorporating lattice fitting, the DGR model \cite{Gubser:2008ny, DeWolfe:2010he} and its extended versions \cite{Critelli:2017oub, Grefa:2021qvt,Cai:2022omk, Zhao:2022uxc, Li:2023mpv, Zhao:2023gur} also construct a family of five-dimension black holes through a non-conformal dilaton potential $V(\phi)$. Therefore we can compare our machine learning model with extended DGR models. The results of $V(\phi)$ and $f(\phi)$, obtained from machine learning $A(z)$ and compared with extended DGR models, are shown in Fig.\ref{comhphi}. It is observed that the results obtained by inputting $A(z)$ are in qualitatively good agreement with those from the extended DGR models using the non-conformal dilaton potential $V(\phi)$, indicating the success of the EMD framework in describing QCD matter.

{\it Phase diagram in $(T, \mu_B)$ plane and the location of CEP: }
The critical temperature at finite $\mu$ can be determined by the minimum of the sound velocity. The phase diagram in the $T-\mu$ plane obtained in the machine learning holographic model is shown in Fig. \ref{cep}. For both 2-flavor and 2+1-flavor systems, the phase transition is crossover at small chemical potentials and first order at large chemical potentials. The CEP for 2-flavor system is located at ($\mu_B^c$=0.46 GeV, $T^c$=0.147 GeV) and for 2+1-flavor system is at ($\mu_B^c$=0.74 GeV, $T^c$=0.094 GeV). The predicted location of CEP for the 2+1-flavor system from this model is very close to recent results from other nonperturbative models, e.g.,  DSE-FRG \cite{Gao:2020qsj}, FRG \cite{Fu:2019hdw} and realistic PNJL model\cite{Li:2018ygx} as well as the extended DGR model in \cite{Critelli:2017oub}. The freeze-out line with corresponding collision energy is also shown in Fig. \ref{cep}. Our predicted CEP is above the freeze-out line, from analysis in \cite{Li:2018ygx}, it might indicate a peak of baryon number fluctuation $\kappa \sigma^2$ appears in the collision energy of $\sqrt{s}\sim 3-5$GeV.

{\it Conclusion and outlook:}
In this work, by using the machine learning method, an analytic holographic QCD model is constructed from the lattice QCD results at zero chemical potential on EoS and baryon number susceptibility.  With machine learning analytic $A(z)$ and $f(z)$, it is straightforward to calculate other quantities. We showed the predicted critical temperatures at vanishing chemical potential and the location of CEP for different systems.  The different locations of CEP in 2-flavor and 2+1-flavor systems reveal that dynamic quarks influence the location of the CEP. Notably, the CEP location in our model for the 2+1-flavor case is close to those from other non-perturbative models, e.g., DSE-FRG \cite{Gao:2020qsj},  FRG \cite{Fu:2019hdw} and realistic PNJL model\cite{Li:2018ygx} as well as the extended DGR model in \cite{Critelli:2017oub}. The consistent results from the machine learning metric and the non-conformal dilaton potential indicate the robustness of the EMD framework in describing QCD matter at finite temperature and chemical potential.

This work represents the first attempt to construct an analytical holographic model using machine learning. This analytical model can give different phase structures for different flavors. We hope that this method will be beneficial for the search of CEP in the QCD phase diagram and help us get a deeper understanding of the hadron spectra within the domain of strong interactions. We aim to incorporate more information into the holographic QCD and construct an even more realistic holographic model in future work with machine learning.

\section*{Acknowledgments}
We thank useful discussion with Danning Li, Zhibin Li, Lingxiao Wang, Yan-Qing Zhao, and Lin Zhang. This work is supported in part by the National Natural Science Foundation of China (NSFC) Grant Nos: 12235016, 12221005, 12147150 and the Strategic Priority Research Program of Chinese Academy of Sciences under Grant No XDB34030000, and the Research Foundation of Education Bureau of Hunan Province, China(Grant No. 21B0402) and the Natural Science Foundation of Hunan Province of China under Grants No.2022JJ40344.

\section*{References}

\bibliography{ref}

\end{document}